\documentclass[12pt]{article}
\usepackage{graphicx}

\begin{document}

\centerline{\bf Superfluid $T_c$ of Helium-3 and its Pressure Dependence}

\vspace{1.3cm}

\centerline{\bf Yatendra S. Jain}

\centerline{Department of Physics}

\centerline{North-Eastern Hill University, Shillong - 793 022, India}

\vspace{1.6cm}

\begin{abstract}

Superfluid $T_c$ of liquid helium-3 and its pressure dependence are calculated 
by using a relation obtained from our macro-orbital microscopic theory.  The 
results agree closely with experiments.  This underlines the accuracy of our 
relation and its potential to provide superfluid $T_c$ of electron fluid in 
widely different superconductors and renders experimental foundation to our 
conclusion ({\it cond-mat/0603784}) related to the basic factors responsible 
for the formation of ({\bf q}, -{\bf q}) bound pairs of fermions and the 
onset of superfluidity in a fermionic system.  Since available experimental 
studies of superconductors pertaining to changes in lattice parameters around 
their superconducting $T_c$ seem to support a link between lattice strain and 
the onset of superconductivity, need for similar studies is emphasized.   

\end{abstract}



\bigskip
email: ysjain@email.com

\vspace{2.6cm}

\noindent
{\bf 1. Introduction}

\bigskip
Liquid $^3He$ has been a subject of extensive theoretical and experimental 
studies [1-5] for the last six decades for several reasons including its 
superfluid behavior and it appears that it will continue to fascinate the 
researchers for many more decades to come.  However, as remarked by Georges 
and Laloux [6] several aspects of even normal state of the liquid at low 
temperatures (LT), {\it viz.}: (i) increase in inertial mass as revealed by 
the experimental values of its LT specific heat, (ii) many fold increase in 
its magnetic susceptibility indicating as if it is at the blink of 
ferromagnetic instability, and (iii) nearly temperature independent low 
compressibility need better understanding.  Two models, {\it viz.}: 
(i) ``almost ferromagnetic'' [7] and  ``almost localized'' [8,9] have been 
extensively tried to account for these aspects.  Identifying that the two 
models are seemingly contradictory, Georges and Laloux [6] propose Mott-stoner 
liquid model.  However, in view of our recently developed microscopic model 
of a {\it system of interacting fermions} (SIF) used to conclude the basic 
foundations of superconductivity [10] both these pictures coexist.  We, 
therefore, have a detailed program to study different aspects of liquid 
$^3He$ ({\it including those listed above}) in the framework of our model 
which is based on the macro-orbital representation of a particle in a 
many body system.  We also use this representation to conclude the long 
awaited microscopic theory a system of interacting bosons such as liquid 
$^4He$ [11].  

\bigskip 
It is evident that the results and inferences of our model [10] can be 
applied to understand the normal and superfluid behavior of liquid $^3He$.  
We note that some of our conclusions to some extent agree with those 
of the well known BCS theory [12] of superconductivity.  For example we find 
that superconductivity is a consequence of {\it bound pairs} of electrons 
moving with equal and opposite momenta ({\bf q}, -{\bf q}).  But in variance 
with BCS theory, the binding of such pairs is basically found [10] to be a 
consequence of the mechanical strain in the lattice forced by the zero-point 
force of electrons arising from their zero point energy; the electrical 
polarization of the lattice emphasized by the BCS model may have its $+ve$ or 
$-ve$ contribution to this binding.  In addition our approach reveals a 
single theoretical framework for the superconductivity of conventional 
as well as high $T_c$ systems and finds that superconductivity can, in 
principle, be observed at temperatures as high as room temperature 
(RT).  It renders mathematically simple formulations and microscopic 
foundations to the well known phenomenological theories ({\it viz.} two 
fluid theory of Landau [13] and $\Psi-$theory of Ginzburg [14]).  It 
concludes that superfluid and superconducting transitions are a kind of 
{\it quantum phase transitions} which, however, occur at a non-zero $T$ due 
to proximity effect of quasi-particle excitations.  Guided by all these 
factors, we use our approach to: (i) estimate the value of superfluid 
$T_c$ of liquid $^3He$ which has been identified to be a difficult 
problem [1], (ii) study its pressure dependence, and (iii) analyze their 
consistency with experiments.  The details of other important properties 
of the liquid would be analyzed in our forthcoming paper(s).    

\bigskip
Theoretical calculations, predicting possible value(s) of $T_c$ of 
superfluid $^3He$, based on 
BCS picture were reported within a year of the publication of the BCS theory.  
While the first few studies [15, 16] indicated that the liquid 
was unlikely to have a superfluid transition, Emery and Sessler [17] 
concluded that a second order transition may occur at a $T$ between 50 to 
100 mK.  However, when the transition was  really observed between 
0.92 to 2.6 mK [18] (depending on the pressure on the liquid), calculations 
by Levin and Valls [19, 20] not only obtained a 
$T_c$ close to these values but also found its pressure dependence closely 
matching with experiments.  Almost similar estimates have been reported 
recently by Rasul and coworkers [21, 22].  Widely different inferences 
and estimated $T_c$ from different theoretical calculations using common 
picture (BCS Theory) as their central idea seem to indicate the complexity 
of estimates and lack of reliability.  On the other hand, the merit of 
our theory [10] lies with the fact that it does not have any scope 
to use different considerations to obtain different $T_c$ which indicates 
its reliability.  In addition the fact that our $T_c$ for superfluid $^3He$ 
its pressure dependence agrees closely with experiments indicates its 
accuracy.

\bigskip
\noindent
{\bf 2  Superfluid $T_c$ and its Pressure Dependence}

\bigskip
Using the universal component ($H_o(N)$, Eqn. 2 of [10]) of the net 
hamiltonian $H(N)$ of a SIF (Eqn. 1 of [10]), such as electron fluid in a 
conductor or liquid $^3He$, we find that its particles below  
$$T_c \approx \epsilon_g/k_B = 
\frac{h^2}{8k_Bmd^2}\frac{\Delta d}{d} \eqno(1)$$

\noindent
assume a state of bound pairs and the system as a whole has a transition to 
its superfluid state [10].  $\epsilon_g$ and $m$ in Eqn. 1, respectively, 
represent the binding energy (or energy gap) and mass of a fermion with (i) 
$d = (V/N)^{1/3)}$ and (ii) $\Delta d =$ an increase in $d$ forced by the 
zero-point force of a fermion occupying its ground state in a cavity 
(size $=d$) formed by neighboring fermions.  It may be noted that for 
electrons in a conductor $d$ in Eqn. 1 represents diameter $d_c$ of the 
channels {\it through which conduction electrons move in the lattice} [10].  
In view of the fact revealed from the experimentally observed specific heat 
values of liquid $^3He$, a particle in an interacting environment of the 
liquid at a $T$ closed to $T_c$ starts behaving like a quasi-particle of 
mass $m^*$, we use 
$$T_c = \frac{h^2}{8k_Bm^*d^2}\frac{\Delta d}{d} \eqno(2)$$

\bigskip
\begin{figure}
\includegraphics [angle = -90, width = 0.9\textwidth]{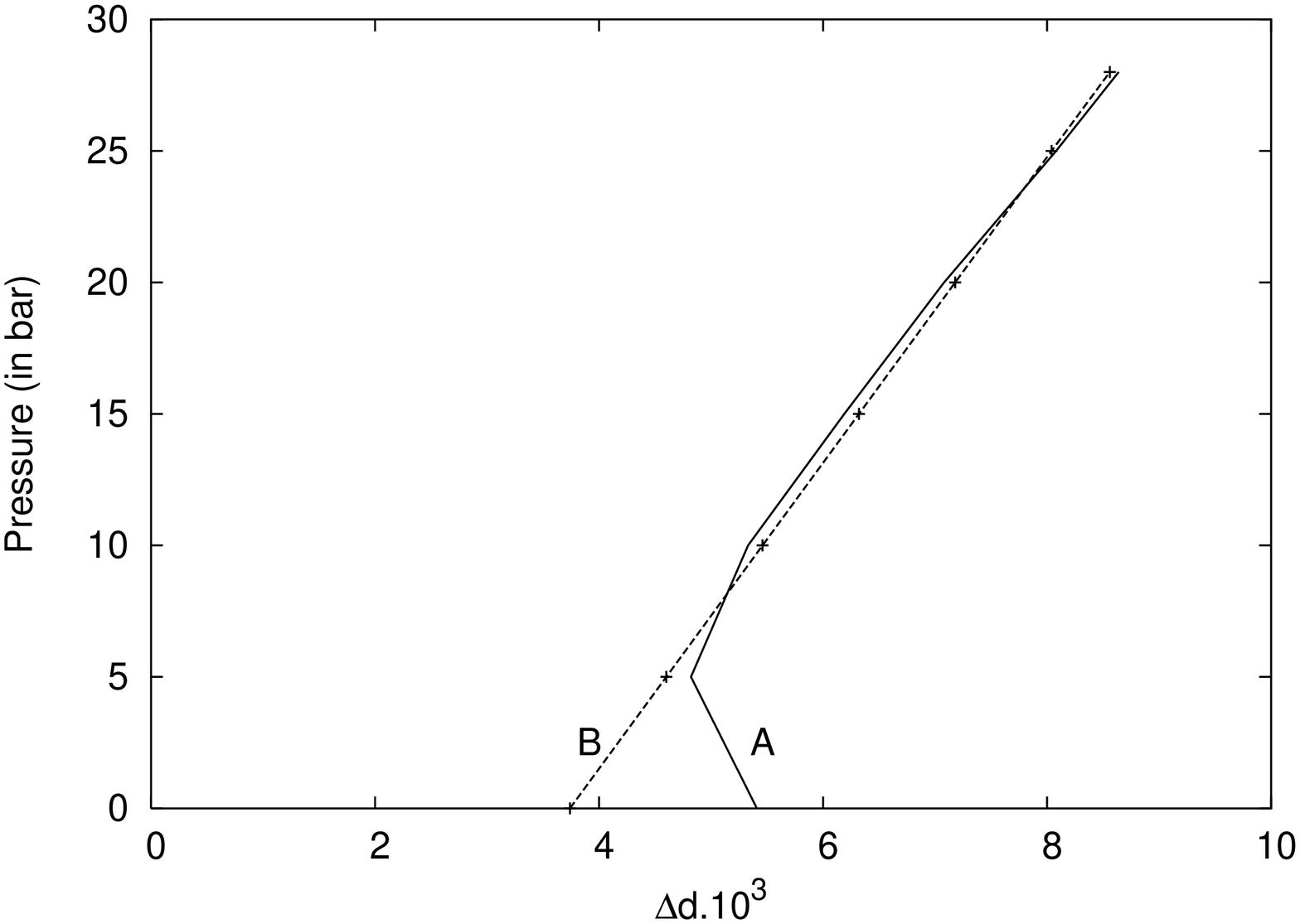}

\bigskip
\noindent
Fig. 1 : Pressure dependence of expansion in $He-He$ bond $\Delta d$
(in \AA).  While Curve A is obtained from the molar volume data of 
Kollar and Vollhardt [24], Curve B represents the linear fit (Eqn. 3) 
after excluding point P=0 at Curve A (see text).  
 
\end{figure}  

\noindent
to obtain $T_c$ of liquid $^3He$ at different pressures.  In this we 
define $\Delta d = d(T=0) - d_{min}$ with $d_{min} = d$ at the point of 
maximum density of the liquid for a chosen pressure.   As shown for the  
simple case of a particle trapped in 1-D box [23], we identify [10] the
zero-point force of a particle occupying its ground state in the cavity 
of neighboring particles as the microscopic reason for the expansion of 
the liquid on cooling below certain $T < T_F$ (Fermi temperature).  We 
determine $d$ and $\Delta d$ by using molar volume of the liquid recently 
reported by Kollar and Vollhardt [24].  However, as indicated by Kollar 
and  Vollhardt [24] themselves and the plot of $\Delta d$ {\it vs.} $T$ 
in Figure 1, their data for $P=0$ seem to have large systematic errors; note 
that $\Delta d$ at $P=0$ falls considerably away from any logical trend in 
which other points can be fitted.  Consequently, we discarded this point 
and obtained a linear fit 
$$P = 5806.15\Delta d - 21.6865   \eqno (3)$$

\noindent
for all other points by using a standard computer software.  In this 
context not only the remaining points seem to fall closely on the line but 
a linear change in $\Delta d$ with increasing $P$ is also expected because 
$\Delta d$ is a kind of strain in $He-He$ bonds.  As such we used Eqn.3 
to obtain $\Delta d$ values for our calculations of $T_c$ at different 
pressures including $P = 0$.  To obtain $m^*$ that enters in Eqn. 2, we 
note that as per our theoretical formulation the quasi-particle excitations 
which contribute to the specific heat of the fermionic system of 
non-interacting particles have $4m$ mass.  Obviously, when the impact of 
interactions is included, we have $4m^*$ as the mass of the quasi-particle 
which, obviously, equals the effective mass ($m^*_{sp}$) that we obtain 
from specific heat data [25].  In other words we use $m^* = m^*_{sp}/4$ in 
Eqn. 2 to obtain our $T_c$ at which the superfluid phase transition is 
expected as per our theory [10].  The $T_c$ values so obtained are tabulated 
with experimental values in Table I and both are plotted in Fig. 2 for their 
comparison.  The fact that our $m^*$ changes from 0.7525$m(^3He)$ to 
1.4233$m(^3He)$ with pressure increasing from 0 to 28.0 bar indicates 
that inter-particle interaction dominated locally by zero-point repulsion 
slowly assumes attractive nature (at $\approx$ 10 bar pressure) with $^3He$ 
atoms having increased electric dipolemoment with increasing pressure.       

\bigskip
\begin{figure}
\includegraphics [angle = -90, width = 0.9\textwidth]{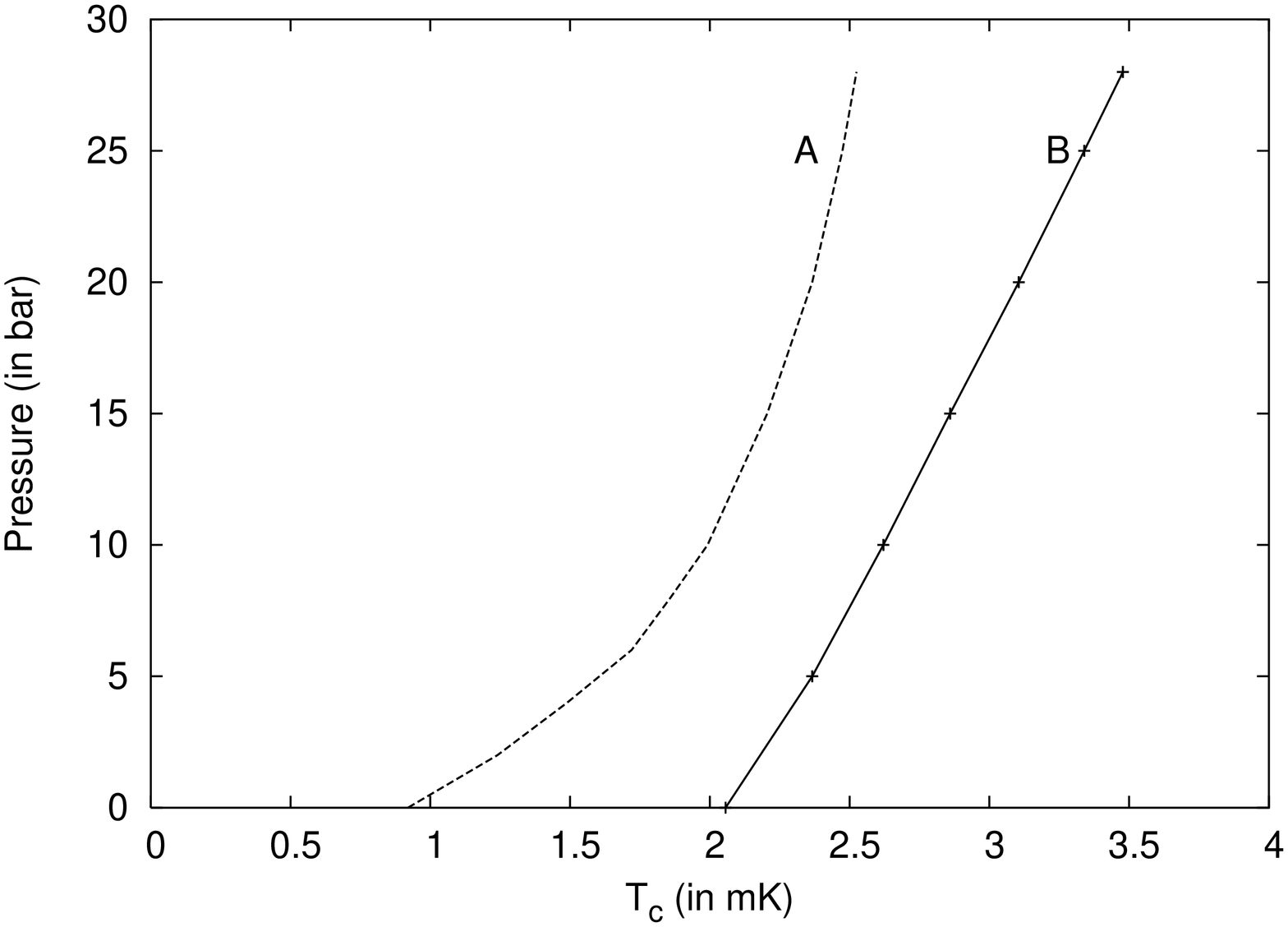}

\bigskip
\noindent
Fig. 2 : Pressure dependence of superfluid $T_c$ of liquid $^3He$. 
Curve A (experimental, {\it cf.} Column 7, Table I) and Curve B (Eqn 2).
 
\end{figure}  

\bigskip
\begin{center}
\noindent
Table I : Calculated and experimental $T_c$ and related data

\bigskip 
\begin{tabular}{ccccccc}\hline\hline

Pressure& $d$&\quad$\Delta d$&\quad $m^*_{sp}$$^+$ &\quad$m^*$ &\quad$T_c(eqn.2)$&\quad $T_c(Exp)$$^{++}$\\

(in bar) & $\AA$ & \quad $\AA$ & \quad $m(^3He)$ &\quad $m(^3He)$ & \quad mK &\quad mK\\
\hline

0.0 & \quad 3.94  & \quad .00374 & \quad 3.010  & \quad 0.7525 & \quad 2.0564  & \quad 0.92 \\
5.0 & \quad 3.78  & \quad .00460 & \quad 3.629  & \quad 0.9073 & \quad 2.3667  & \quad 1.60 \\
10.0 & \quad 3.69  & \quad.00546 & \quad 4.183  & \quad 1.0458 & \quad 2.6208  & \quad 1.99 \\
15.0 & \quad 3.63  & \quad.00632 & \quad 4.670  & \quad 1.1675 & \quad 2.8598  & \quad 2.21 \\
20.0 & \quad 3.58  & \quad.00718 & \quad 5.084  & \quad 1.2710 & \quad 3.1057  & \quad 2.37 \\
25.0 & \quad 3.54  & \quad.00804 & \quad 5.472  & \quad 1.3680 & \quad 3.3399  & \quad 2.47 \\
28.0 & \quad 3.52  & \quad.00856 & \quad 5.693  & \quad 1.4233 & \quad 3.4782  & \quad 2.52 \\

\hline
\end{tabular}
\end{center}
$^+$ obtained from graphical plots of $m^*_{sp}$ values [25],  $^{++}$Zero 
pressure value from [26] and others from [25].

\newpage

\noindent
{\bf 3. Discussion}

\bigskip
The BCS model, basically formulated to explain superconductivity of 
conventional superconductors, has been used to understand 
superfluidity and related aspects of liquid $^3He$ because the electron 
fluid in conductors and liquid $^3He$ are closely identical SIF; of course 
suitable modifications ({\it e.g.}, fermions participating in Cooper pairing 
in the latter case have $p-$state not $s-$state) compatible with the model 
are adopted.  This paper uses the same considerations to apply the basic 
foundations of superconductivity [10] revealed from our non-conventional 
theoretical framework which emphasizes mechanical strain ({\it in the 
crystalline lattice of a superconductor or in inter-atomic bonds in case of 
liquid $^3He$ type} SIF) as the main source of ({\bf q}, -{\bf q}) bound pair 
formation.  As established in [23, 27], such strain is 
{\it a basic consequence of zero-point force arising from the wave particle 
duality} and it ought to be present whenever a particle occupies its ground 
state in a box or cavity of its neighboring particles or in a channel through 
which it is free to move.  While electrons in superconductors create strain 
in the lattice structure of the channels through which they move [10], 
a $He$ atom creates this strain in $He-He$ bonds which it makes with its 
neighboring atoms [11].  The experimental fact that liquid $^3He$ as well as 
liquid $^4He$ show $-ve$ thermal expansion at $T \approx T_o$ 
({\it the temperature equivalent of the ground state energy of $He$ atom in 
a cavity of neighboring atoms}) confirms the presence of strain in $He-He$ 
bonds.  Evidently, our theoretical estimate of superfluid $T_c$ of liquid 
$^3He$ and its pressure dependence ({\it cf.}, Table 1 and Figure 2), which 
have close agreement with experiments [25,26], undoubtedly prove the accuracy 
of Eqn. 2 and conclusions of [23 and 27].  It also demonstrates the potential 
of our theory [10] to predict the superfluid $T_c$ of a SIF which has been a 
difficult task in the framework of conventional BCS theory [1].  In other 
words Eqn. 2 can be used to estimate the superconducting $T_c$ of widely 
different superconductors (including {\it high $T_c$ superconductors}) if 
accurate values of $d$, $\Delta d$ and $m^*$ are known.   Several experimental 
studies [{\it e.g.}, 28-33] indicate that the occurrence of lattice strain 
or related effects such as negative expansion of lattice, hardening of 
lattice, anomalous or anisotropic change in lattice parameters, {\it etc.} 
around superconducting $T_c$ are common aspects of superconductors.  This 
naturally supports our inference [10] regarding the relation between lattice 
strain and bound pair formation.  However, the effect is not seen to be 
as clean as in liquid $^3He$ because an electron in a superconductor not 
only interacts with other electrons but also to the ions or atoms which 
decide their lattice arrangement through complex inter-particle 
interactions.  Naturally, an accurate prediction of superconducting $T_c$ 
from Eqn.2 depends on the accuracy of the experimentally measured 
$\Delta d/d$, $d$ and $m^*$ for a chosen superconductor.  In view of 
Eqn.2, $T_c$ increases with increase in $\Delta d/d$ and decrease in $d$ 
and $m^*$ and, depending on the values of these parameters, superfluid 
transition in a SIF can, in principle, occur at any temperature.  This 
is corroborated by the facts that: (i) an atomic nucleus exhibits nucleon 
superfluidity at a $T$ much higher than even room temperature because 
nucleon-nucleon $d$ is found to be about $10^{-5}$ times shorter than 
$^3He-^3He$ distance which implies that the typical $T_c$ should be as 
high as $10^7$K (the mass of a nucleon and $^3He$ atom having same 
order of magnitude), and (ii) a typical superconducting $T_c$ falls 
around 10 K ($\approx 10^3$ times the superfluid $T_c$ of liquid $^3He$) 
because $m_e$ (mass of electron) is about 6000 times smaller than 
$m(^3He)$ (inter-electron distance or the channel size being nearly 
equal to $^3He - ^3He$ distance).

\bigskip
We note that Eqn.2 has been obtained by analyzing the universal component,  
$H_o(N) = H(N) - V'(N)$, of the net hamiltonian $H(N)$ ({\it cf.}, Eqn. 1 
of [10]) of a SIF with $V'(N)$ representing the sum of all non-central 
potentials including spin-spin interactions; as such it considers only bare 
fermion-fermion central forces.  Evidently, our estimates of $T_c$ of 
superfluid $^3He$ and its pressure dependence (Table 1 and Figure 2) exclude 
the contributions from $V'(N)$ (sum of interactions such as spin-spin 
interactions, electron-phonon interaction induced by electric polarization 
of the lattice, {\it etc.}) and possibly for this reason our estimates 
are about two times higher than experimental values.  In view of these facts 
our estimates not only establish that the ``mechanical strain'' forced by 
zero-point force is the basic cause of bound pair formation in a SIF but also 
indicates that $V'(N)$ perturbations could be responsible for the supression 
of $T_c$ below our estimates.  We hope that this would be supported by 
studies related to the impact of these perturbations on $T_c$ and its 
pressure dependence.  Heiselberg {\it et. al.} [34] have summarized the 
important inferences of studies related to the impact of induced 
interactions such as BCS type attraction on the $T_c$ deduced from bare 
fermion-fermion interaction.  They identify that such interactions in 
liquid $^3He$ are responsible for the ABM state to be energetically 
more favorable than BM state, while in neutron matter they suppress the 
superfluid gap significantly.  The effect has been studied in dilute spin 
1/2 Fermi gas by Gorkov and Melik-Barkhurdarov [35] who find that $T_c$ 
obtained from bare inter-particle interactions gets suppressed by a 
factor $(4e)^{1/3} \approx 2.2$.  Evidently, all such effects of  
$V'(N)$ interactions (not included in deriving Eqn. 2) can be trusted 
for reducing the difference of our values of $T_c$ with experiments 
(Table 1 and Figure 2).  We plan to examine these effects in our future 
course of studies.  Interestingly, similar results of pressure dependent 
$T_c$ reported in [19-22] are also about two times higher than 
experimental values but it appears that these studies leave no factor(s) 
which could help in getting better agreement with experiments.                 
        
\bigskip
Finally, it may be mentioned that we have limited information about the 
thermal expansion of superconductors [36] around $T_c$ while the importance 
of its detailed and accurate measurements has been emphasized [37] soon after 
the discovery of high $T_c$ superconductors.  The observation of negative 
lattice expansion, anisotropic thermal expansion, change in hardness, 
{\it etc.} around $T_c$ in a number of superconducting systems [28-33, and 
38-40] not only re-emphasizes the importance of such studies but also 
indicates a relation of this effect with the onset of superfluidity in 
fermionic systems which naturally corroborates its mechanism as concluded by 
our theory [10].          

\bigskip
\noindent
{\bf 4. Conclusion}  

\bigskip
The paper uses a relation obtained from our recently reported theoretical 
model [10] to estimate superfluid $T_c$ of $^3He$ and its pressure 
dependence.  The close agreement between our estimates and experimental 
results indicates the accuracy of our model and the microscopic mechanism 
of superfluidity in a SIF like liquid $^3He$ and electron fluid in 
widely different superconductors.  As suggested in [28], we also believe 
that accurate measurements of different aspects related to modifications in 
lattice structure, {\it viz.} thermal expansion, changes in lattice 
parameters, hardening, change in sound velocity, {\it etc.} around 
superconducting $T_c$ of widely different superconductors would be of 
great help in establishing the role of mechanical strain in the lattice as 
a basic component of the microscopic mechanism of superfluidity of different 
SIF and we hope that these would support our theory [10].  In this context 
it may be noted that liquid $^3He$ and liquid $^4He$ which do not 
have various complexities of electron fluid in conductors exhibit $-ve$ 
thermal expansion around superfluid $T_c$ as predicted by their respective 
microscopic theories [10] and [11] based on our macro-orbital approach.  In 
addition it is significant that our approach has no space for subjective 
considerations which provide widely different estimates of $T_c$ as one may 
see with different $T_c$ values (0 to 100 mK) [15-17, 19-22] estimated by 
using only one model ({\it the BCS picture}) for superfluid transition in 
liquid $^3He$.


\begin{thebibliography}{??}

\bibitem[1]{Kn:gnus} 
A. J. Leggett, Rev. Mod. Phys. {\bf 76}, 142 (2004); 
 \bibitem[2]{Kn:gnus}
D.M. Lee, Rev. Mod. Phys. {\bf 69}, 645 (1997).
\bibitem[3]{Kn:gnus} 
D. Vollhardt and P. Wolfle, {\it The Superfluid Phases of Helium -3}, Taylor and Francis, London (1990).
\bibitem[4]{Kn:gnus} 
G.E. Volovik, {\it Exotic Properties of Superfluid $^3He$}, World Scientific, Singapore (1992).  
\bibitem[5]{Kn:gnus} 
G.E. Volovik, {\it The Universe in a Helium Droplet}, Clarendon Press, Oxford (2003),
\bibitem[6]{Kn:gnus} 
A. Georges and L. Laloux, {\it Normal Helium 3 : a Mott Stoner liquid}, arXiv:cond-mat/9610076 (1996).
\bibitem[7]{Kn:gnus} 
K. Levin and O.T. Valls, Phys. Rep. {\bf 98}, 1 (1983).
\bibitem[8]{Kn:gnus} 
P.W. Anderson and W.F. Brinkman in {\it The Helium Liquids}, J.G.M. 
Armitage and I.E. Farqhar, Eds., Academic, New York (1975). 
\bibitem[9]{Kn:gnus} 
D. Vollhardt, Rev. Mod. Phys. {\bf 56}, 99 (1984).
\bibitem[10]{Kn:gnus} 
Y.S. Jain, {\it Basic Foundations of the Microscopic Theory of Superconductivity}, arXiv:cond-mat/0603784 (2006).
\bibitem[11]{Kn:gnus} 
Y.S. Jain, {\it Macro-orbitals and Microscopic Theory of a System of Interacting Bosons}, arXiv:cond-mat/0606571 (2006).
\bibitem[12]{Kn:gnus} 
J. Bardeen, L.N. Cooper and Schrieffer, Phys. Rev. {\bf 106}, 162 (1957).
\bibitem [13]{Kn:gnus} 
L.D. Landau, J. Phys. (USSR) {\bf 5}, 71 (1941); English translation 
published in {\it Helium 4} by Z.M. Galasiewicz, Pergamon Press, Oxford 
(1971), pp 191-233.
\bibitem[14]{Kn:gnus} 
V.N. Ginzburg, Rev. Mod. Phys. {\bf 76}, 981 (2004).
\bibitem[15]{Kn:gnus} 
N.N. Bogoliubov, Doklady Akad. Nauk (U.S.S.R.) {\bf 119}, 1 (1958); 
Sov. Phys. Doklady {\bf 3}, 292 (1958).
\bibitem[16]{Kn:gnus} 
L.N. Cooper, R.L. Mills and A.M. Sessler, 
Phys. Rev. {\bf 114}, 1377 (1959).
\bibitem[17]{Kn:gnus} 
V.J. Emery and A.M. Sessler, Phys. Rev. {\bf 119}, 43 (1960).
\bibitem[18]{Kn:gnus} 
D.D. Osheroff, R.C. Richardson and D.M. Lee, Phys. Rev. Lett.  
{\bf 28}, 885 (1972). 
\bibitem[19]{Kn:gnus} 
K. Levin and O.T. Valls, Phys. Rev. B {\bf 17}, 191 (1978).
\bibitem[20]{Kn:gnus} 
K. Levin and O.T. Valls, Phys. Rev. B {\bf 20}, 105 (1979).
\bibitem[21]{Kn:gnus} 
J.B. Rasul, T.C. Li and H. Beck, Phys. Rev. B {\bf 39}, 4191 (1989).
\bibitem[22]{Kn:gnus} 
J.B. Rasul, Phys. Rev. B {\bf 45}, 4191 (1992).
\bibitem[23]{Kn:gnus} 
Y.S. Jain, {\it Untouched Aspects of the Wave Mechanics of a Particle in 1-D 
Box}, arXiv:quant-ph/0606009 (2006).
\bibitem[24]{Kn:gnus}
M. Kollar and D. Vollhardt, Phys. Rev. B {\bf 61}, 15347 (2000).
\bibitem[25]{Kn:gnus} 
J.C. Wheatley in {\it The Helium Liquids}, J.G.M. 
Armitage and I.E. Farqhar, Eds., Academic, New York (1975). 
\bibitem[26]{Kn:gnus} 
C. Enss and S. Hunklinger, {\it Low Temperature Physics} Springer-Verlag, 
Berlin 2005.  
\bibitem[27]{Kn:gnus} 
Y.S. Jain, {\it Wave Mechanics of Two Hard core Particles in 1-D 
Box}, arXiv:quant-ph/0603233 (2006);  Cent. Euro. J Phys. 
{\bf 2}, 709 (2004). 
\bibitem[28]{Kn:gnus} 
A. J. Mills and K.M. Rabe, Phys. Rev. B {\bf 38}, 8908 (1988).
\bibitem[29]{Kn:gnus} 
C. Meingast, O. Kraut, T. Wolf, H. Wuhl, A. Erb and G. Muller-Vogt, 
Phys. Rev. Lett. {\bf 67}, 1634 (1991).
\bibitem[30]{Kn:gnus} 
G. J. Burkhart and C. Meingast, Phys. Rev. B {\bf 54}, R6865 (1996).
\bibitem[31]{Kn:gnus} 
J. Kortus, I.I. Mazin, K.D. Belashchenko, V.P. Antropov and L.L. Boyer, 
Phys. Rev. Lett. {\bf 86}, 4656 (2001).
\bibitem[32]{Kn:gnus} 
J.M. An and W.E. Pickett, Phys. Rev. Lett. {\bf 86}, 4366 (2001).
\bibitem[33]{Kn:gnus} 
T. Yildirim et al, Phys. Rev. Lett. {\bf 87}, 037001 (2001).
\bibitem[34]{Kn:gnus} 
H. Heiselberg, C.J. Pethick, H. Smith and L. Viverit, Phys. Rev. Lett. 
{\bf 85}, 2418 (2000).
\bibitem[35]{Kn:gnus} 
L.P. Gorkov and T.K. Melik-Barkhudarov, Sov. Phys. JETP {\bf 13}, 1018 (1961).
\bibitem[36]{Kn:gnus}
T.H.K. Barron, J.G. Collins and G.K. White, Adv. Phys. {\bf 29}, 609 (1980).
\bibitem[37]{Kn:gnus}
A. de Visser {\it et al} Phys. Rev. B {\bf 41}, 7304 (1990).
\bibitem[38]{Kn:gnus}
T. Takeuchi, H. Shishido, S. Ikeda, R. Settai, Y. Haga and Y. Onuki, 
J. Phys.: Condens Matter {\bf 14}, L261 (2002).
\bibitem[39]{Kn:gnus}
A.C. Mclaughlin, F. Sher and J.P. Attfield, Nature {\bf 436}, 829(2005).
\bibitem[40]{Kn:gnus} 
J.D. Jorgensen, D.G. Hinks, P.G. Radaelli, W.I.F. David, R.M. Ibberson, 
{\it Large Anisotropic Thermal Expansion Anomaly near the Superconducting 
Transition TEmperature in MgB$_2$}, arXiv:cond-mat/0205486 (2002). 

\end{thebibliography}
\end{document}